%% file: main.tex
\documentclass[reprint, superscriptaddress, onecolumn, aps]{revtex4-2} 
\input{article_config}
\input{commands}

\begin{document}

\author{David S. Kammer} \email[Corresponding author: ]{dkammer@ethz.ch} \affiliation{Institute for Building Materials, ETH Zurich, Switzerland} 
\author{Gregory C. McLaskey} \affiliation{School of Civil and Environmental Engineering, Cornell, University, Ithaca, NY, USA}

\author{Rachel E. Abercrombie} \affiliation{Boston University, Boston, MA, USA}
\author{Jean-Paul Ampuero} \affiliation{Université Côte d’Azur, IRD, CNRS, Observatoire de la Côte d’Azur, Géoazur, France}
\author{Camilla Cattania} \affiliation{Department of Earth, Atmospheric, and Planetary Sciences, Massachusetts Institute of Technology, Cambridge, MA, USA}
\author{Massimo Cocco} \affiliation{Istituto Nazionale di Geofisica e Vulcanologia, Rome, Italy}
\author{Luca Dal Zilio} \affiliation{Institute of Geophysics, ETH Zurich, Switzerland}
\author{Georg Dresen} \affiliation{Helmholtz Centre Potsdam, GFZ German Research Centre for Geosciences, Potsdam, Germany}
\author{Alice-Agnes Gabriel} \affiliation{Scripps Institution of Oceanography, UCSD, La Jolla, USA}%
\author{Chun-Yu Ke} \affiliation{Department of Geosciences, The Pennsylvania State University, University Park, PA, 16802, USA} 
\author{Chris Marone} \affiliation{La Sapienza Università di Roma, P.le Aldo Moro 5, 00185 Roma, Italia} \affiliation{Department of Geosciences, The Pennsylvania State University, University Park, PA, 16802, USA}
\author{Paul A. Selvadurai} \affiliation{Swiss Seismological Service, ETH Zurich, Switzerland}
\author{Elisa Tinti} \affiliation{La Sapienza Università di Roma, P.le Aldo Moro 5, 00185 Roma, Italia} 


\title{Energy dissipation in earthquakes}

\begin{abstract}
Earthquakes are rupture-like processes that propagate along tectonic faults and cause seismic waves. The propagation speed and final area of the rupture, which determine an earthquake's potential impact, are directly related to the nature and quantity of the energy dissipation involved in the rupture process. Here we present the challenges associated with defining and measuring the energy dissipation in laboratory and natural earthquakes across many scales. We discuss the importance and implications of distinguishing between energy dissipation that occurs close to and far behind the rupture tip and we identify open scientific questions related to a consistent modeling framework for earthquake physics that extends beyond classical Linear Elastic Fracture Mechanics.
\end{abstract}

\maketitle

Earthquakes are one of the most damaging natural hazards facing humankind. Improvements in understanding the fundamental physics of earthquakes could have dramatic consequences for our ability to plan and react to catastrophic earthquakes in densely populated areas. Seismological observations show that earthquakes comprise a rupture front propagating along a fault and leaving behind slip and stress drop, which is a form of fracture propagation. Thus the field of fracture mechanics has played a fundamental role in shaping what we know about earthquake physics. Classical models describe an earthquake as a shear crack and define, for example, the relationship between earthquake rupture area, propagation speed, and the spectral characteristics of radiated seismic waves~\citep{Brune1970,boatwright1980spectral,madariaga1976}. While the overall energy budget that compares states before and after an earthquake has a well-established theoretical basis~\citep{kostrov1974seismic}, key aspects of the instantaneous energy balance governing the behavior of the earthquake rupture remain poorly understood. Fracture mechanics theory predicts that rupture growth is a balancing act involving three components: 1) energy dissipated to extend the crack, either by creating new surface area or generating frictional heat, 2) energy radiated as seismic waves, and 3) release of stored elastic energy from the surrounding rock. This view has been confirmed by a broad range of laboratory experiments and codified in the theory of Linear Elastic Fracture Mechanics (LEFM)~\citep{freund1998dynamic}. However, the complexity of earthquake faults far exceeds that of typical laboratory setups, raising significant questions about the applicability and predictive power of LEFM in natural conditions~\citep{kostrov1988principles,rudnicki1980fracture}. One of our goals here is to paint a picture of the state of the art in understanding earthquake rupture and in particular the extent to which LEFM can further our understanding of the earthquake energy budget in cases where the fault zone has finite width and rupture propagation involves branching, off-fault fracturing and other processes that go beyond the simple assumptions underlying LEFM. 

While there is transformative potential in extending knowledge and connections between the fields of fracture mechanics and earthquake physics, there are several key impediments. These include: 1) a lack of fundamental understanding of how various dissipative processes at different spatial and temporal scales contribute to the mechanics of earthquakes, 2) extreme discrepancies (of many orders of magnitude) between values of fracture energy measured in laboratory experiments~\citep{cocco2023fracture} and inferred from natural earthquakes, and 3) vastly different terminology between the communities. In this Perspective, we aim to review the mechanics and energy dissipation in earthquake ruptures, identify translations of terminology, and discuss the capabilities and limitations of current observations and measurement techniques and how they affect the observed discrepancies. Finally, we propose a path forward in the form of key outstanding questions, clear scientific objectives for future work, and suggestions to overcome the limitations of LEFM as applied to earthquake faulting. 

\addfig{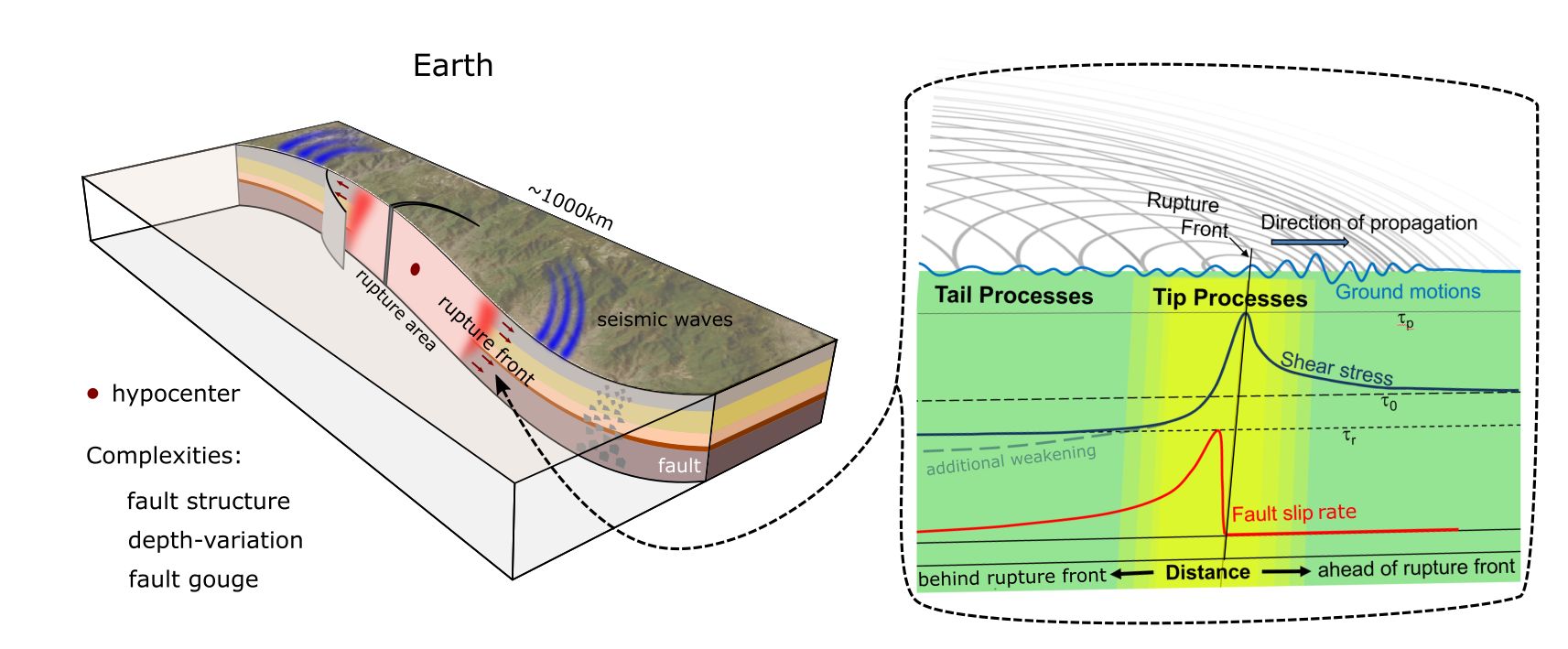}{Schematic of an earthquake with its rupture front and seismic waves. (left) Earthquake fault complexity includes fault geometry, depth variation in geological units, wear and gouge formation within the fault zone, and fault branching. The red zone near the rupture front indicates regions with high fault slip rates. Note seismic waves (blue) radiate from the fault zone. (right) The evolution of the fault zone shear stress (blue) and slip rate (red) are shown for the region around the rupture tip and toward the hypocenter where the fault has already slipped. Seismic waves are indicated by gray lines and the associated ground motions are illustrated by a bright blue line. Processes within the rupture tip and in the tail behind the tip, so-called tip-and-tail processes, are a major focus of this Perspective.
}{0.75}

\section{Fundamentals of theoretical earthquake mechanics} \label{sec:fundamentals}

The mechanics of earthquakes are complex and arguably equally challenging to measure as to theoretically describe. The goal of a theoretical earthquake model is to describe the essential processes with viable equations and tools that allow field observation to be interpreted. Hence, it builds on common observations, which show that earthquake ruptures begin in a localized region of a fault known as the hypocenter (see Fig.~\ref{fig:earthquake_complexity_v4_01}), which is the location where initial shear stresses ($\shearstress$) are sufficient to overcome frictional strength and the fault motion begins to accelerate. The initial stress level is one of the most difficult parameters to constrain. It can be spatially variable and it can have a large impact on rupture style and velocity. 
Unlike static frameworks, such as slip-tendency analysis popular in structural geology, faults can be stressed well below strength almost everywhere and yet rupture spontaneously: only a small portion of the fault needs to reach its strength to nucleate an earthquake. 
For ordinary earthquakes with fast rupture speeds the hypocenter is the location where seismic waves are first radiated. From there, earthquake rupture expands along the fault, causing the fault surfaces to begin to slip. The transition region between slipping and unslipped sections of the fault is called the rupture front (see Figs.~\ref{fig:earthquake_complexity_v4_01}). For fast ruptures, moving at speeds of several $\mathrm{km/s}$, the slip rate of fault surfaces accelerates from below 1 \textmu m/s to above 1 m/s over timescales of less than a second. Ahead of the rupture front, the shear stress on the fault increases in what is called a dynamic stress concentration. At the rupture tip, a rapid transition occurs. The slip velocity increases as the shear stress drops rapidly to a dynamic level $\dynamicfriction$ that is below the initial level $\shearstress$. This drop in stress causes a release of stored strain energy, which drives the rupture. Simultaneously, part of this energy is dissipated through various processes including fracture of the surrounding rock, comminution (the production of rock powders), heating and possibly melting of rocks. The remaining energy is radiated as seismic waves that transport kinetic energy far from the source and cause ground shaking. Eventually, the rupture ceases to grow, and all sections of the earthquake rupture area arrest (or evolve to a very slow and long quasi-static front of postseismic slip). This can occur because continued slip necessitates a disproportionately large amount of energy dissipation or because the rupture front propagates into unfavorably stressed regions (\textit{i.e.}, $\shearstress < \dynamicfriction$).

\vspace{5mm}
\noindent\fbox{%
\parbox{\textwidth}{%
\textbf{Box 1: Fundamentals of Linear Elastic Fracture Mechanics (LEFM)}

Linear Elastic Fracture Mechanics (LEFM) is a theoretical framework to describe crack growth, originally developed with a focus on opening (Mode I) cracks~\citep{griffith1921vi,zehnder2012fracture,broberg1999cracks}. The question of whether a crack grows or not is reduced to a comparison of two states (see Fig.~\ref{fig:lefm_v01_02}): the current state with crack (half-)length $\crackhalflength$ and an incremental state with crack length $\crackhalflength + \crackhalflengthincrement$. 
The following assumptions are made:
\begin{enumerate}
    \item the material surrounding the crack has predominantly linear elastic behavior, 
    \item dissipation is localized at the crack tip, 
    \item the crack surface is traction-free. 
    \end{enumerate}
One can then express the comparison of states in terms of energies stating that the crack grows if $\staticenergyreleaserate > \totalfractureenergy$, where $\staticenergyreleaserate$ is the static energy release rate and $\totalfractureenergy$ the fracture energy~\citep{irwin1956onset,williams1957on}. The static energy release rate is the drop in the elastic energy $\Pi$ stored in the medium per increment of crack length $\staticenergyreleaserate = - \mathrm{d}\Pi/\crackhalflengthincrement \approx -\left(\Pi_2 - \Pi_1\right)/\crackhalflengthincrement$. The fracture energy is the energy dissipated in the process of breaking the material, per unit of crack length growth. 
This comparison of energies is only possible because the above-mentioned assumptions guarantee a ``separation of scale'' between the global quantity $\staticenergyreleaserate$ driving the crack, and the local quantity $\totalfractureenergy$ resisting crack growth. The key aspect of this approach is that the specific physical processes of how energy is dissipated (\textit{e.g.}, as decohesion, plastic work) are irrelevant as long as they are localized at the crack tip, satisfying assumption 2. They are all lumped into the parameter $\totalfractureenergy$. 

Note that assumption 2 leads to singular stresses at the crack tip. To avoid this unphysical phenomenon the dissipation is often smeared out in a still-localized-enough ``process zone'' of size $\processzone \ll \crackhalflength$ -- an approach known as ``small-scale yielding'' and commonly implemented via cohesive zone models~\citep{ida1972cohesive,palmer1973growth,ohnaka2003constitutive}. While the precise limit for small-scale yielding remains unknown, a process zone with $\processzone \approx 0.4\crackhalflength$ may, under some circumstances, still be enough localized \citep[see pp. 142-143 in][]{zehnder2012fracture}.   

LEFM also predicts the speed at which the crack grows~\citep{freund1998dynamic} through an energy balance, $\energyreleaserate = \totalfractureenergy$, where the dynamic energy release rate $\energyreleaserate \leq \staticenergyreleaserate$ accounts for kinetic energy being radiated away from the rupture tip. Assuming time-invariant loading (\textit{i.e.}, no wave reflections) and 2D or circular 3D configurations, the dynamic energy release rate can be approximated by $\energyreleaserate \approx (1-\rupturespeed/\Rayleighwavespeed) \staticenergyreleaserate$, where $\rupturespeed$ is the rupture speed and $\Rayleighwavespeed$ the material Rayleigh wave speed. 
}%
}
\addfig{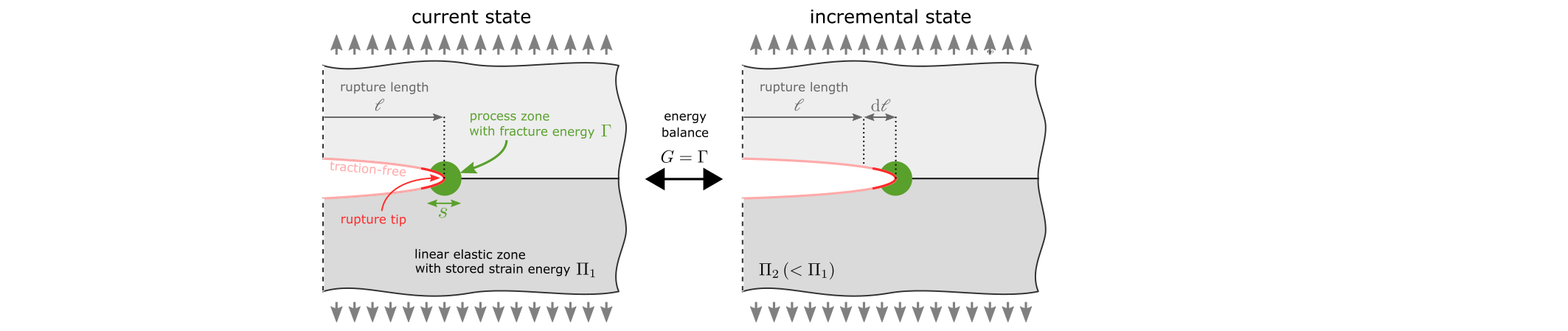}{\textbf{(Figure for Box 1):} Schematic of the fundamental principles for crack growth in Linear Elastic Fracture Mechanics.}{0.99}

\subsection{Earthquakes as a rupture process described by fracture mechanics} \label{sec:fundamentals:LEFM}

The earthquake rupture process of propagation and arrest shares many features of a crack propagating through a solid material. Thus, the theoretical framework of LEFM, as summarized in Box~1, has been adapted to describe the mechanics of earthquake rupture \cite[\textit{e.g.},][]{scholz2019mechanics,palmer1973growth,ida1972cohesive}. Specifically, under suitable assumptions, LEFM provides an energy balance that governs rupture growth: 
\begin{equation}
    \energyreleaserate = \totalfractureenergy ~,
    \label{eq:energybalance}
\end{equation}
where the terms are defined in Box~1. 
 LEFM allows rupture speed prediction based on a few assumptions given in Box~1
\begin{equation}
    \rupturespeed \approx \left( 1 - \frac{\totalfractureenergy}{\staticenergyreleaserate} \right) \Rayleighwavespeed ~.
    \label{eq:equationofmotion}
\end{equation}
However, the application of LEFM to earthquake ruptures requires a few key adaptions from its classical form. For instance, LEFM assumes a traction-free crack behind the rupture front (Box~1). While valid for opening cracks (mode I), this is clearly not valid for surfaces in frictional contact. Moreover, a broad range of earthquake source properties and frictional behaviors are well described by rate-and-state friction laws \cite[\textit{e.g.},][]{Marone1998,Leeman2016}. Despite the assumption of a traction-free surface after the rupture front passes being simplified, it allows treating dynamic friction and hence the residual shear stresses of shear cracks (modes II and III) behind the rupture front as time-independent and approximately constant. That is, one assumes that dynamic friction reaches a constant residual level $\residualfriction$ for slip larger than a characteristic slip distance $\charslip$. While this is inconsistent with expectations for complex natural faults, it provides a starting place for the application of LEFM and an opportunity to investigate the earthquake energy budget and rupture dynamics.

In the attempt to measure energy dissipation, the work done by the frictional stresses above the minimal stress through slip, commonly known as the ``breakdown work'' $\breakdownwork$, is measured/computed (see Box~2). If residual friction is constant and $\charslip$ is small enough, then the breakdown work is spatially localized in the process zone near the rupture tip. This would ensure the above-mentioned separation of scale that allows the application of LEFM because the energy dissipation associated with fracture propagation is determined entirely within the crack tip region, and therefore independent of other processes on the rest of the rupture surface. In this particular case, the breakdown work is exactly equal to the associated fracture energy, \textit{i.e.}, $\totalfractureenergy = \breakdownwork$ (assuming that there are no other dissipative processes). 

\vspace{5mm}
\noindent\fbox{%
\parbox{\textwidth}{%
\textbf{Box 2: The breakdown work}

The breakdown work~\citep{Tinti2005,Cocco2006} is the measurable portion of the frictional work density, which, when integrated on the fault surface, gives an estimate of the irreversible part of the total strain energy change which does not go into radiated energy~\citep{kostrov1988principles,cocco2023fracture}. \cite{Tinti2005} defined the breakdown work $\breakdownwork$ as the excess of work over the minimum shear stress achieved during slip $\tau_\mathrm{min}$:
\begin{equation}
    \breakdownwork = \int_0^{t_b} \left( \tau(t) - \tau_\mathrm{min} \right) \dot \delta (t) dt = \int_0^{d_c} \left(\tau (\delta) - \tau_\mathrm{min} \right) d\delta
    \label{eq:breakdownwork}
\end{equation}
where $\dot \delta(t)$ is the slip rate, $\tau(t)$ the shear stress, and $t_b$ is the time at which $\tau_\mathrm{min}$ and the critical slip distance $d_c$ are reached.
}%
}
\vspace{5mm}

Some important comments:
\begin{enumerate}
    \item \textbf{breakdown work $\neq$ fracture energy}: breakdown work $\breakdownwork$ and associated fracture energy are not \emph{generally} equal, as commonly assumed in some literature. The breakdown work is only equal to the fracture energy under specific conditions, \textit{i.e.}, when there is separation of scale (see Box~1). The precise limit of separation of scale remains unknown. Since only the localized part of the breakdown work is part of the fracture energy, $\totalfractureenergy \leq \breakdownwork$ (in the absence of other dissipative processes).
    \item \textbf{fracture energy and physical processes}: the fracture energy $\totalfractureenergy$ governing crack growth in Eqs.~\ref{eq:energybalance}\&\ref{eq:equationofmotion} is the cumulative quantity that may include the effects of many processes such as work done by frictional weakening, plastic dissipation, off-fault damage, pulverization within the fault zone, and others, \textit{i.e.}, $\totalfractureenergy = \sum \Gamma_{\mathrm{processes}}$. However, $\totalfractureenergy$ only includes the part of the energy that is localized near the crack tip (on- or off-fault) to comply with separation of scale and to be consistent with LEFM.
    \item \textbf{fracture energy variability}: due to its multi-physical origin, the fracture energy may vary spatially along the fault, and change with rupture speed. Hence, it may not be known a-priori.
    \item \textbf{non-localized heat}: the so-called ``frictional heat'', \textit{e.g.}, $\heatwork = \int_0^\averageSlip \residualfriction~\mathrm{d}\delta$, which is the work of $\residualfriction$ on the fault, is not included in $\totalfractureenergy$ as it is \emph{not} localized. Consequently, the energy due to the residual stress is also excluded from $\energyreleaserate$, and the Griffith energy balance (Eq.~\ref{eq:energybalance}).
\end{enumerate}

\subsection{Applicability of LEFM to laboratory ruptures versus natural earthquakes} \label{sec:funadmental:applicability}

Frictional stick-slip events or fracture propagation on pre-existing surfaces represent the laboratory equivalent of earthquakes. While earthquakes generated in the laboratory (so-called ``labquakes'') share many features of natural earthquakes~\citep{Cebry2022} on tectonic faults, the vast differences in scale raise important questions that include the application of LEFM to labquakes and earthquakes on natural faults. Yet, recent works provide a useful starting point because they include quantitative predictions of rupture speed~\citep{svetlizky2017brittle,kammer2018equation} and arrest~\citep{kammer2015linear,bayart2016fracture,ke2018rupture} for labquakes. These and similar experiments also measure the fracture energy $\totalfractureenergy$ of labquakes with local dynamic shear stress measurements~\citep{svetlizky2014classical,bayart2018rupture,kammer2019fracture,xu2019robust} and/or the stress-versus-slip relation~\citep{okubo1981fracture,okubo1984effects}. Such experiments require a sample that is large compared to the process zone size and a critical length scale for rupture nucleation. For fault normal stresses of $1\text{\textendash}10~\mathrm{MPa}$ this requires meter-scale rock samples~\citep{kammer2019fracture,xu2019robust} or $20\text{\textendash}30~\mathrm{cm}$ sized samples composed of glassy polymer such as PMMA \cite[\textit{e.g.},][]{svetlizky2014classical,rubino2017understanding}. Dynamic rupture can also be studied on smaller samples, at higher normal stress levels ($50\text{\textendash}150~\mathrm{MPa}$), if arrays of sensors are installed on the sample, inside a pressure vessel~\citep{passelegue2013sub} and for cases where the fault zone contains sufficient wear material~\citep{Shreedharan2021EPSL}. To infer rupture-related quantities, it is important to measure stress evolution on or near the fault as a rupture front propagates past the sensor location as opposed to sample-wide averages.

LEFM has also been applied to tectonic faults to examine physical processes such as aseismic slip, occurring naturally or by fluid injection \citep[\textit{e.g.},][]{hawthorne2013laterally,garagash2021fracture,dublanchet2019fluid,galis2017induced,dalzilio2022hydro}, the statistical properties of small earthquakes \citep[\textit{e.g.},][]{dublanchet2018dynamics,dublanchet2021dual,cattania2019crack}, the frequency-magnitude distribution \citep[\textit{e.g.},][]{dempsey2016forecasting,cattania2019complex}, earthquake nucleation \citep[\textit{e.g.},][]{rubin2005earthquake,ida1972cohesive,cattania2023source}, and for rupture propagation and arrest \citep[\textit{e.g.},][]{weng2019dynamics}. 
These results demonstrate that LEFM, even with its strong simplifying assumptions, is a powerful concept to describe the fundamental mechanics of earthquakes and faults. 

In summary, the simplifying assumptions of LEFM appear to be valid for large-scale laboratory experiments where LEFM quantitatively describes rupture velocity and arrest. However, laboratory experiments differ from natural faults in several ways that must be accounted for to understand the limitations of LEFM and to develop appropriate extensions of that theory. First, the magnitude range of labquakes is relatively limited, which impedes a precise determination of earthquake scaling properties. Furthermore, laboratory experiments are often conducted at low-stress levels ($\sim 5~\mathrm{MPa}$) compared to the unconfined strength of the rock or polymer samples. This limits off-fault damage or inelastic deformation that may strongly affect the rupture process and the overall energy dissipation. The experiments also typically employ simple fault geometries, while tectonic faults are much more complex. Under some conditions, the complexities of geological faults can be lumped into a single tip-localized parameter $\totalfractureenergy$; however, for other cases the framework of LEFM requires modification. A key question is if energy dissipation (aside from frictional heat) on natural faults with all its complexities (\textit{e.g.}, weakening, off-fault inelasticity) truly is localized in the vicinity of the rupture tip, which would guarantee separation of scale and applicability of LEFM. Should this not be the case, is it enough if ``most'' of the energy is dissipated in a localized manner? The implications of these questions are important as the answers determine the extent to which LEFM can be applied to earthquakes, in its current or modified forms.

\section{Tip or tail: spatiotemporal energy dissipation in earthquakes} \label{sec:tip_or_tail}

Tectonic faulting complexity involves simultaneous dissipative processes during an earthquake, such as fracturing, comminution, heating, and possibly rock melting. These processes depend on the fault slip rate and the thickness of the shearing zone within the fault. Mechanisms such as flash heating \cite[\textit{e.g.},][]{Rice2006}, melt lubrication \cite[\textit{e.g.},][]{Nielsen2008}, thermal pressurization \cite[\textit{e.g.},][]{viesca2015ubiquitous}, acoustic fluidization~\citep{Melosh1996}, elastohydrodynamic lubrication~\citep{Brodsky2001,dalzilio2022hydro}, off-fault deformation incurred during slow~\citep{Rudnicki1975} or fast rupture~\citep{Andrews2005,BenZion2005,Rice2005,Bhat2007,gabriel2013source} are among the various processes that have been proposed to explain energy dissipation during earthquakes.

These dissipative processes may influence the mechanics of earthquakes in different ways~\citep{benzion2022synthesis,cocco2023fracture}. However, building on previous literature \citep[][among others]{paglialunga2022scale,brantut2019stability}, we propose a conceptual picture (see Fig.~\ref{fig:earthquake_complexity_v4_01}) that distinguishes between the following key processes:
\begin{enumerate}
    \item \emph{Tip processes} dissipate energy near the rupture front and therefore contribute to the earthquake fracture energy that is equivalent to $\totalfractureenergy$ utilized in LEFM. The rupture tip region is characterized by intense slip accelerations ($> 100~\mathrm{m/s}^2$) and high slip velocities ($> 1~\mathrm{m/s}$), but because it is highly transient, the associated slip is typically a relatively small fraction of the total coseismic slip.
    \item \emph{Tail processes} occur behind the rupture tip where slip acceleration is much lower. However, slip velocities may remain relatively high ($\sim 1~\mathrm{m/s}$) in the wake of the rupture tip, especially for crack-like ruptures compared to pulse-like ruptures. Therefore, the slip accumulated in the rupture \emph{tail} can be large if the rupture continues long enough. 
\end{enumerate}
 
We note that a sharp boundary between the tip and tail processes likely does not exist and that some dissipative processes are affected by both tip and tail \citep[\textit{e.g.},][]{cornelio2022determination}. The tip and tail terminology is not limited to interface processes (\textit{i.e.}, fault processes) but may also include dissipation in the bulk material (\textit{i.e.}, host rocks), consistent with the initial formulation of LEFM. 
The ``tip'' and ``tail'' terminology becomes specifically useful when discussing how different dissipative processes may affect different aspects of earthquake rupture propagation and arrest. For example, flash heating may be a weakening mechanism that is active as a tip process. In contrast, thermal pressurization will likely only occur as a tail process after sufficient slip has occurred \citep[\textit{e.g.},][]{lambert2020rupture}. A particularly valuable aspect of \emph{tip and tail} is that it highlights a fundamental question: does energy dissipation in the \emph{tail} contribute to the energy balance for rupture extension or is crack propagation determined entirely at the rupture tip? LEFM predicts that ruptures will extend following $\energyreleaserate = \totalfractureenergy$, but for \emph{tail} processes that are sufficiently remote from the rupture tip, it is not clear how they contribute to the energy needed for crack extension $\totalfractureenergy$.

Laboratory experiments offer valuable insights into this problem, albeit with notable limitations. Most small-scale experiments, for instance, cannot achieve slip acceleration that is fast enough to fully emulate the loading conditions of a dynamic rupture front (\textit{i.e.}, tip processes), while large-scale rupture experiments do not exhibit enough slip for tail processes to become dominant. It is far from trivial to set up laboratory experiments capable of reproducing the slip values, velocities, and accelerations under realistic loading conditions representative of a propagating earthquake rupture. These challenges can be addressed with numerical simulations, which allow us to assess tip processes under non-trivial friction conditions or assess contributions by other dissipative mechanisms, under limited conditions \citep[\textit{e.g.},][]{andrews1976rupture,Andrews2005}. Aside from limitations of laboratory and theoretical work, we also note that the theoretical definition of where the tip ends and the tail starts is not well defined and is likely model-dependent \citep[]{cornelio2022determination} and hence requires further investigation.

Finally, how do the tip and tail processes influence the mechanics of earthquakes? This is one of the key open questions in earthquake physics and is at the center of this Perspective. Without tail processes, LEFM shows, as outlined in Sec~\ref{sec:fundamentals}, that the dissipative energy in the rupture tip controls rupture speed and arrest. Whether this is equally true for systems with significant tail processes remains to be shown. Recent results~\citep{barras2020emergence,paglialunga2022scale,weng2022integrated} suggest that the tip processes dictate the rupture growth even in the presence of non-negligible tail dissipation, so it may be reasonable to speculate that the tip processes primarily control rupture propagation and arrest. However, the tail may become important when the earthquake propagates slowly, possibly during a slow arrest, propagates as multiple fronts, as a self-healing slip pulse~\citep{heaton1990evidence}, or in multiple sliding episodes across rough fault surfaces\citep[\textit{e.g.},][]{fang2013additional}, when considering the complex prestress state~\citep{ampuero2008cracks} or when considering how earthquakes prepare the fault for subsequent events. Here again, numerical simulations provide a tool to systematically study the link between tip and tail processes and the mechanics of earthquake ruptures.

In summary, defining tip and tail processes and how various dissipative processes contribute to them is crucial to a better understanding of how earthquakes propagate, arrest, and prepare the fault for subsequent events. The size of the yielding zone near the tip of a propagating rupture front is also an open question, which affects the values of inferred fracture energy as we will discuss in the next section. Laboratory experiments and numerical simulations, best in synergistic combination, may provide crucial insight into these processes but require further development. Finally, the proposed framework needs to be applied to natural earthquakes but this requires a precise understanding of how field observations are linked to these tip and tail processes, which is another important open question.

\section{Observations of energy dissipation in natural, laboratory, and simulated earthquakes}


Given these theoretical considerations, we explore and compare observations of energy dissipation in both labquakes and tectonic earthquakes, with a focus on what this reveals about applying LEFM in earthquake physics. Estimates of $\totalfractureenergy$ and the total energy dissipation vary widely for labquakes, tectonic earthquakes, and numerical models of earthquake rupture. For instance, estimates by \citet{Abercrombie2005}, which are calculated from a combination of seismically derived parameters (see Box~3), have suggested that the \emph{average} energy dissipation in natural earthquakes ranges across multiple orders of magnitude from $10^{2}$ to $10^{7}~\mathrm{J/m}^2$. A compilation of seismologically inferred energy dissipation extends this range to $10^{-2}\text{\textendash}10^{8}~\mathrm{J/m}^2$~\citep{viesca2015ubiquitous,cocco2023fracture}, but these estimates are all highly model-dependent and subject to large, and potentially systematic, uncertainties~\citep{Abercrombie2021}. Pseudo-dynamic earthquake modeling that infers shear stress evolution based on slip history \citep[\textit{e.g.},][]{mai2014srcmod} yields estimates of $~\mathrm{MJ/m}^2$ for magnitudes $M$ larger than $5$~\citep{Tinti2005} but commonly shows large variability along the fault plane~\citep{Tinti2005,Bouchon1997,Causse2014,Ide1997}. Near-fault observations are used to infer constitutive parameters, such as the critical slip distance ~\citep{Kaneko2017}, which would imply large values of breakdown work or \emph{average} energy dissipation. Other approaches based on dynamic models~\citep{Tinti2021,Gallovic2019,Premus2022} that simulate spontaneous dynamic rupture and employ a frictional constitutive law yield estimates of total energy dissipation that range from 1 to 10 $~\mathrm{MJ/m}^2$. 

\vspace{5mm}
\noindent\fbox{%
\parbox{\textwidth}{%
\textbf{Box 3: Seismologically derived earthquake parameters}

\citet{Abercrombie2005} proposed a parameter $\Gprime$, here denoted $\seismologicalbreakdownwork$ to avoid confusion with energy release rate:
\begin{equation}
    \seismologicalbreakdownwork = \frac{\averageSlip}{2}\left(\stressdrop-\frac{2{\mu}E_\mathrm{R}}{M_0}\right)~,
\end{equation}
where $\averageSlip$ is the average slip over the fault plane, $\stressdrop$ the average stress drop, $\mu$ the rock shear modulus, $E_\mathrm{R}$ the radiated energy, and $M_0$ the seismic moment. $\seismologicalbreakdownwork$ is theoretically equal to the breakdown work as long as the final stress $\overline{\tau}^\mathrm{E}_\mathrm{f}$ is equal to the residual sliding strength of the fault $\tau_\mathrm{r}$. This approach is entirely based on seismologically derived parameters and can be based on rupture averages from simple source models, or derived from finite-fault modeling with spatially and temporally varying slip. 
The most reliable seismologically derived parameter is the seismic moment 
\begin{equation}
\seismicmoment = \mu A \averageSlip~,
\end{equation}
where $A$ is the rupture area. In finite-fault models, the stress drop can be determined from the spatially varying slip, but for smaller earthquakes, it is typically determined by assuming a simple circular source model~\citep{Brune1970}: 
\begin{equation}
\stressdrop = \frac{7}{16}\frac{\seismicmoment}{r^3}~.
\end{equation}
For circular ruptures, $r$ can be estimated from the corner frequency of the spectrum of teleseismic waves ($f_0$), or the reciprocal of the pulse duration in time domain modeling,
\begin{equation}
r = \frac{k}{\beta}{f_0}~,
\end{equation}
where $k$ is a geometrical constant that varies widely for commonly used source models~\citep{Brune1970,madariaga1976,kaneko_shearer_2015}.
These seismologically determined source parameters are subject to large systematic and random uncertainties and should be interpreted and modeled with extreme caution~\citep{Abercrombie2021}.
}%
}%
\vspace{5mm}

Similar estimates of energy dissipation from acoustic emission spectra of labquakes yield values of $10^{-6}\text{\textendash}1~\mathrm{J/m}^2$~\citep{Selvadurai2019}. In more traditional shear fracture experiments of intact rock, however, the fracture energy has been measured in the range $10^{3}\text{\textendash}10^{4}~\mathrm{J/m}^2$ at the $100~\mathrm{MPa}$ pressures expected in much of the seismogenic crust~\citep{Wawersik1971,Wong1985,Wong1982,Wong1986,Lockner1991}. This is the energy required to form a fault in intact rock and is supposed to be orders of magnitude higher than the fracture energy required to rupture an existing tectonic fault. Energy dissipation estimated from friction experiments on rough surfaces that are flat at long wavelength yield estimates of $10^{-1}\text{\textendash}10^{1}~\mathrm{J/m}^2$~\citep{okubo1981fracture,okubo1984effects,kammer2019fracture,xu2019robust}. Other experiments where surfaces experience large slip and concentrated shear heating show continued weakening of the interface up to $1~\mathrm{m}$ of slip~\citep{nielsen2016scaling,di2011fault}, which has been interpreted as energy dissipation up to $1~\mathrm{MJ/m^2}$. Still other data come from mining-induced earthquakes where the faults intersect working faces. The fraction of energy dissipated during an M2.1 mining event was estimated to be about $1\text{\textendash}9\%$ of the total energy released~\citep{olgaard1983microstructure,McGarr1974}. More recently \cite{chester2005fracture} arrived at a similar fraction of $<1\%$ of the total energy based on fault gouge analysis, which corresponds to an average dissipated energy per event of roughly $0.5~\mathrm{MJ/m}^2$.

The interpretation of the broad range of values inferred for energy dissipation (per unit area) requires careful analysis, as some may correspond to (tip-localized) fracture energy and others correspond to energy dissipation within a broader region that is not localized and should therefore not be included in fracture energy. For example, quasi-static laboratory experiments, including rotary shear and other experiments with modest slip acceleration ($5~\mathrm{m/s}^2$) produce conditions appropriate for tail processes, not tip processes. Thus, under these conditions, the inferred dissipation does not correspond to the fracture energy that affects the rupture tip $\breakdownwork \neq \breakdownfractureenergy$. Spectral seismological estimates \citep[\textit{e.g.}, compilations of ][]{Abercrombie2005,viesca2015ubiquitous} use information from the entire earthquake rupture area, rather than from just the propagating rupture front, and can therefore also include `tail' dissipation mechanisms. 

It has also been shown~\citep{Guatteri2000} that the resolution of shear stress evolution inferred from pseudodynamic modeling \citep[\textit{e.g.},][]{Tinti2005} is limited by the bandwidth of the input data, either from the band limits imposed on ground motions or from smoothing operators used to regularize the kinematic finite-fault inversions. However, integral quantities such as fracture energy or breakdown work are less affected by bandwidth limitations and thus they are considered more reliable measures. Scale dependence of the physical processes governing dynamic weakening might also explain both the broad range of values and the scaling of energy dissipation with slip \citep[see][and references therein]{cocco2023fracture}. Given the tip-and-tail separation outlined above, future research is needed to determine the values of $\totalfractureenergy$ and the total energy dissipation retrieved at different scales and use different techniques to evaluate both laboratory data and natural earthquakes. We emphasize here that the wide range of inferred values for fracture energy and total energy dissipation do not agree with predictions from LEFM, and this discrepancy will motivate future research.

Several recent works have discussed the increase of fracture energy and breakdown work with total slip or earthquake size. A coherent interpretation of this scaling is still lacking, which implies some caution. However, we note that this scaling is observed in seismological estimates from both natural earthquakes~\citep{Abercrombie2005} and labquakes~\citep{Selvadurai2019} as well as in numerical modeling studies, as shown in \cite{cocco2023fracture}. On the other hand, \cite{Ke2022}  proposed a numerical model to suggest that scaling of seismologically estimated energy dissipation with slip can result from stress overshoot, rather than a true increase of fracture energy with rupture size and fault slip. While overshoot cannot be used to explain the scaling reported from pseudo-dynamic modeling \cite{Tinti2005} that work raises important questions about how to reconcile the huge range of fracture energy measurements and the earthquake energy budget. This highlights the importance of further research to reconcile the constant fracture energy assumption of LEFM with current practices used to perform dynamic modeling of natural earthquakes and ground motion predictions.

\section{Conclusion \& Outlook}

This Perspective discusses the earthquake energy budget and the potential of Linear Elastic Fracture Mechanics (LEFM) theory to describe earthquake rupture in the laboratory and nature. The key condition for LEFM applicability is that energy dissipation other than heat generation must be localized at the rupture tip -- a condition that is commonly satisfied in large-scale laboratory experiments but may not be fully met for tectonic faults with all of their complexity. This raises important questions about how to consistently and correctly describe the energy dissipation of natural earthquakes. We suggest distinguishing between \emph{tip} processes that account for localized dissipation and \emph{tail} processes that occur further away from the rupture tip. In this framework, tip processes govern earthquake rupture extension and propagation while tail processes are more important for other measures of earthquake energy dissipation and the global energy budget.
We also highlight the large range of measured or inferred energy dissipation from labquakes and earthquakes across many orders of magnitude, and the possibility that this could result from comparing localized with non-localized dissipation. 

While the proposed tip-versus-tail perspective provides a useful approach to discussing energy dissipation in earthquakes, it also opens important scientific questions that are to be addressed in future research. For instance, the boundary between the tip and tail, \textit{i.e.}, the localization of energy dissipation, is neither well defined nor known. Experiments and field observations with improved sensing are needed to measure the contributions to tip and tail energy dissipation. Here it is important to note that any physical process may contribute to dissipation in the tip and the tail concurrently, and hence separating contributions to each is required. It is also important to use precise and consistent terminology to avoid misinterpretation of data. Specifically, only rupture-tip energy dissipation should be termed ``fracture energy''; and when there is no proof that energy dissipation is local to the rupture tip more general terms, such as ``breakdown work'', should be used.

Another important open question concerns the effect of significant tail processes on earthquake propagation and arrest mechanisms. Do these processes affect the energy balance (Eq.~\ref{eq:energybalance}) and, hence the rupture speed? Here, numerical simulations are particularly useful to systematically explore and isolate these effects and to update the fracture theory for the description of rupture growth in the presence of tail processes. Such simulations could also provide a tool to determine the link between fault properties (tip and tail processes) and averaged global observations as inferred from seismological data and hence support the correct interpretation of earthquake energy dissipation across scales.

In conclusion, as a community, we need to synergistically combine field observations, laboratory experiments, and numerical simulations to determine the degree of rupture-tip localization of various energy dissipative processes, and the effect of non-localized dissipation on rupture mechanics to build a consistent model for earthquake physics.

\bibliography{references}

\section*{Acknowledgements}
We thank the organizers of the Workshop on Earthquake Dynamics: Mechanical Work and Fracture Energy, which was the starting point for this work.

\section*{Author contributions}
D.S.K. coordinated the writing of the manuscript. D.S.K. and G.C.M. wrote the initial draft. All authors contributed to the editing of the manuscript.

\section*{Competing interests}
All authors declare no competing interests.

\end{document}

%% file: article_config.tex
\usepackage{amsmath,amssymb,amsfonts}
\usepackage{graphicx}
\usepackage{dcolumn}
\usepackage{bm}
\usepackage{hyperref}
\usepackage{color,soul}
\usepackage{comment}
\usepackage{booktabs}
\usepackage[table]{xcolor}

\usepackage[sort&compress]{natbib}

\usepackage[colorinlistoftodos]{todonotes}
\setuptodonotes{inline}

\date{\today}

\newcommand{\addfig}[3]{\begin{figure}[ht] \centering \includegraphics[width=#3\textwidth]{#1} \caption{#2} \label{fig:#1} \end{figure}} 

%% file: commands.tex
\newcommand{\rupturespeed}{C_\mathrm{f}}
\newcommand{\Rayleighwavespeed}{C_\mathrm{R}}

\newcommand{\residualfriction}{\tau_\mathrm{r}}
\newcommand{\dynamicfriction}{\tau_\mathrm{d}}

\newcommand{\shearstress}{\tau_0}

\newcommand{\energyreleaserate}{G}
\newcommand{\staticenergyreleaserate}{\energyreleaserate_0}

\newcommand{\fractureenergy}{\Gamma}
\newcommand{\totalfractureenergy}{\fractureenergy_\mathrm{tot}}
\newcommand{\breakdownfractureenergy}{\fractureenergy_\mathrm{b}}
\newcommand{\plasticfractureenergy}{\fractureenergy_\mathrm{pl}}
\newcommand{\offfaultfractureenergy}{\fractureenergy_\mathrm{off}}
\newcommand{\heatwork}{W_\mathrm{H}}

\newcommand{\breakdownwork}{W_\mathrm{b}}
\newcommand{\seismologicalbreakdownwork}{W'_\mathrm{b}}
\newcommand{\plasticwork}{W_\mathrm{pl}}
\newcommand{\offfaultwork}{W_\mathrm{off}}
\newcommand{\crackhalflength}{\ell}
\newcommand{\crackhalflengthincrement}{\mathrm{d}\ell}
\newcommand{\processzone}{s}
\newcommand{\charslip}{d_\mathrm{c}}
\newcommand{\averageSlip}{D}
\newcommand{\Gprime}{G'}

\newcommand{\seismicmoment}{M_\mathrm{0}}
\newcommand{\stressdrop}{\Delta\sigma}